# Silicon Nanoridge Array Waveguides for Nonlinear and Sensing Applications

**Matthew W. Puckett[1,3*], Rajat Sharma[1,*], Felipe Vallini[1], Shiva Shahin[1], Faraz Monifi[1], Peter N. Barrina[1], Soroush Mehravar[2], Khanh Kieu[2], and Yeshaiahu Fainman[1]**

[1]*Department of Electrical & Computer Engineering, University of California, San Diego, 9500 Gilman Dr., La Jolla, California 92023, USA6, USA*
[2]*College of Optical Sciences, University of Arizona, 1630 E. University Blvd., Tucson, Arizona 85721, USA*
*\*equal contributors*
[3]*mwpucket@ucsd.edu*

**Abstract:** We fabricate and characterize waveguides composed of closely spaced and longitudinally oriented silicon ridges etched into silicon-on-insulator wafers. Through both guided mode and bulk measurements, we demonstrate that the patterning of silicon waveguides on such a deeply subwavelength scale is desirable for nonlinear and sensing applications alike. The proposed waveguide geometry simultaneously exhibits comparable propagation loss to similar schemes proposed in literature, an enhanced effective third-order nonlinear susceptibility, and high sensitivity to perturbations in its environment.

©2014 Optical Society of America

**OCIS codes:** (160.4236) Nanomaterials; (280.4788) Optical sensing and sensors; (190.4390) Nonlinear optics, integrated optics; (070.7345) Wave propagation.

## 1. Introduction

Over the past two decades, integrated silicon photonics has advanced to the brink of large-scale commercialization and become one of the most promising candidates in overcoming the bottlenecks of conventional CMOS circuits [1-3]. More recently, chip-scale photonic devices have also been used in such applications as biological and environmental sensing [4,5]. Perhaps the most critical effect which allows silicon waveguides to exhibit improved nonlinear coefficients, response times, and sensitivities is the confinement of electromagnetic fields to a small spatial area [6-9]. In addition to drastically increasing the spatial power density of the supported modes, the reduction of a waveguide's cross-section also raises its surface-area-to-volume ratio, allowing otherwise negligible phenomena at material interfaces to become strong in comparison to bulk effects [10,11]. As silicon photonics continues to evolve, it will be important to design new waveguide topologies capable of exploiting these surface phenomena without consequently spoiling other favorable properties.

Slot waveguides show promise as one means by which this may be achieved, and have previously been used to realize high-efficiency electro-optic modulators and wavemixers [12-14]. Whereas conventional waveguides confine the optical mode to a high-index core, slot waveguides exploit the discontinuities of electric fields across interfaces to trap the mode within the lower-index slot. Sub-wavelength gratings (SWGs), which modify a waveguide's cross-section along the direction of propagation, are also appealing for various applications due to their high sensitivities to changes in the waveguide's cladding material [15]. A natural integration of these two waveguide configurations, designed with the intent of simultaneously leveraging both of their advantageous properties, would consist of many thin, co-propagating ridges of guiding material, as shown in Fig. 1. Such a waveguide, patterned over dimensions drastically smaller than the wavelength of light, would preserve the field localization achieved by the slot waveguide while simultaneously allowing the guided mode's effective optical properties to be easily tailored. In this manuscript, we consider the linear and nonlinear characteristics of the so-called nanoridge array (NRA) waveguide and assess its applicability to several photonic applications.

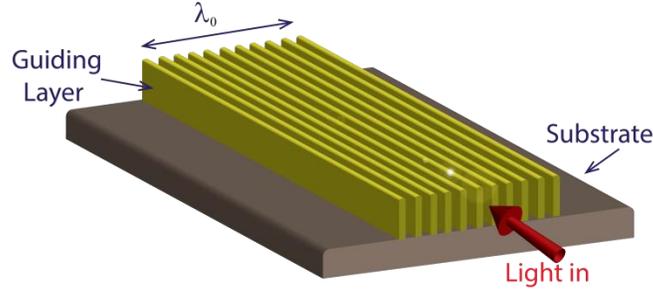

Fig. 1. Illustration showing the size of the proposed waveguide relative to the optical carrier's approximate free-space wavelength.

## 2. Basic properties of silicon nanoridge array (NRA) waveguides

In order for an NRA structure to function as a waveguide, it must operate above the cutoff condition for either the fundamental TE- or TM-like mode. However, for applications such as sensing, it is desirable to operate as close to these same cutoff conditions as possible in order to maximize sensitivity. To more explicitly consider this inherent trade-off between confinement and local field enhancement, we consider two cases in which an NRA waveguide is either (1) left unclad, or (2) clad with aluminum oxide. The waveguides in both cases are composed of 500 nm-tall silicon ridges, and differ only in terms of the material used to clad them. We use the finite element method (FEM) software COMSOL to model each case, and the results yielded by this analysis are shown in Fig. 2 [16].

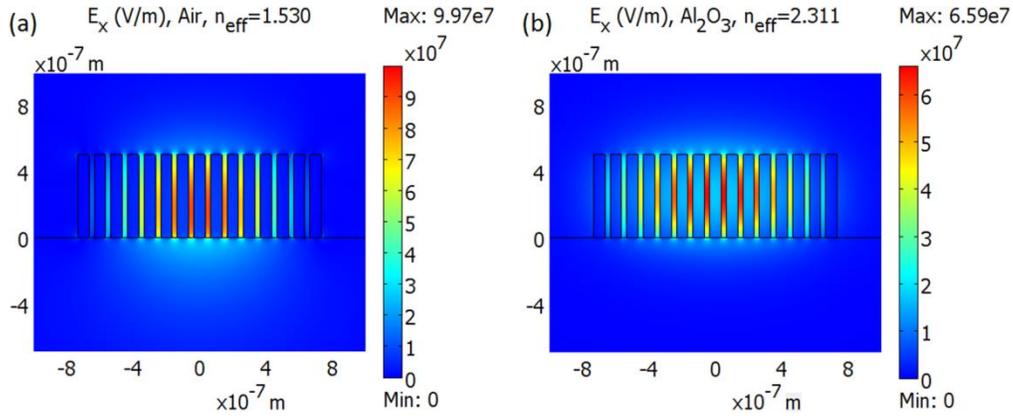

Fig. 2. FEM models showing the x-component of the electric field for the TE-like mode supported by the (a) unclad and (b) aluminum oxide clad NRA waveguide.

The TE-like mode of the unclad waveguide, shown in Fig. 2a, has large electric field discontinuities at each vertical interface. In addition to interacting strongly with the waveguide's densely integrated silicon-air interfaces, this mode is anticipated to respond strongly to any changes in the refractive index of the space between the ridges. These characteristics are obviously desirable for nonlinear and sensing applications, but they come at the expense of confinement, leading to a large modal area as well as leakage into the waveguide's substrate. In comparison, the mode supported by the clad waveguide, plotted in Fig. 2b, shows comparatively less interaction with the cladding material. This may be beneficial, however, because the mode will remain well-confined for smaller waveguide dimensions, whereas the unclad waveguide's TE-like mode would more quickly reach cutoff. Thus, depending on the intended application of the NRA waveguide, it is critical to determine the proper values of such parameters as the ridge width, the pitch between adjacent ridges, the total number of ridges, the height of the waveguide, and the cladding material.

A second complication faced by the proposed waveguide geometry involves the potential difficulty in exciting its optical modes. Coupling light into an NRA waveguide abruptly from a conventional waveguide is anticipated to be highly inefficient due to the large discrepancy between the two guided modes' electric field profiles. Higher coupling efficiencies are predicted, however, if the conventional waveguide is adiabatically tapered into the NRA configuration as illustrated in Fig. 3a. To verify this prediction we have modeled both cases using the finite-difference time-domain software Lumerical [17], the top-view of which is shown as an inset in Fig. 3b. The 3-D simulated structure consisted of a 220 nm tall NRA waveguide clad with 150 nm of aluminum oxide and a micron of silicon dioxide. Our simulations, the results of which are illustrated in Fig. 3b, clearly indicate that the tapered case couples light into the TE-like mode of the NRA waveguide more efficiently than the abrupt transition, and additionally inhibits reflection back into the counter-propagating mode. This analysis offers some evidence that fabricating and characterizing an NRA waveguide may be feasible, and additionally highlights several of the underlying complexities necessarily entailed by the use of such an atypical waveguide geometry.

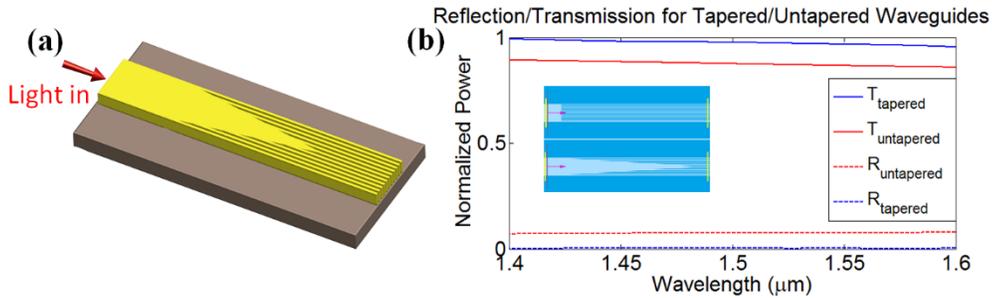

Fig. 3. a) Schematic showing the simulated structure (not including the top clad of aluminum oxide and silicon dioxide) b) Normalized transmission and reflection spectra for the tapered and untampered cases (inset: top view of the simulated structures in Lumerical FDTD).

## 3. Fabrication

The proposed waveguide design, shown in Fig. 4a, was fabricated using commercially available SOI wafers with a 220 nm-thick silicon device layer and a 3 μm-thick buried oxide (BOX) layer. The wafers were spin-coated with the negative electron beam resist hydrogen silsesquioxane (HSQ), and the waveguide pattern was written using a Vistec EBPG 5200 Electron Beam Writer. Subsequently, an Oxford Plasmalab 100 Reactive Ion Etcher was used to remove the unwritten portion of the silicon device layer. The gas flow rates during the etching process were 20 sccm for $SF_6$ and 55 sccm for $C_4F_8$, and the chamber pressure was maintained at 15 mTorr. These conditions were chosen in order to maintain smooth waveguide sidewalls, and to minimize undercut. After etching, the waveguides were clad with a 150 nm-thick layer of thermally deposited $Al_2O_3$ in a Beneq TFS200 Atomic Layer Depositor (ALD) at a temperature of 200° C, and finally an additional 2 μm-thick layer of $SiO_2$ was deposited onto the samples using an Oxford Plasmalab 100 Plasma-Enhanced Chemical Vapor Depositor (PECVD).

The resulting waveguide cross-section, shown in Fig. 4b, exhibited several deviations from the target dimensions. Most prominently, the silicon ridge width was seen to vary with height, reaching a minimum value of approximately 35 nm at the ridge's midpoint.

Additionally, air voids were observed at regular intervals, likely due to clogging between adjacent ridges during the ALD step. Otherwise, the fabricated waveguides were in good agreement with the proposed geometry, and were predicted to support TE- and TM-like modes as shown in Fig. 4c and Fig. 4d, respectively.

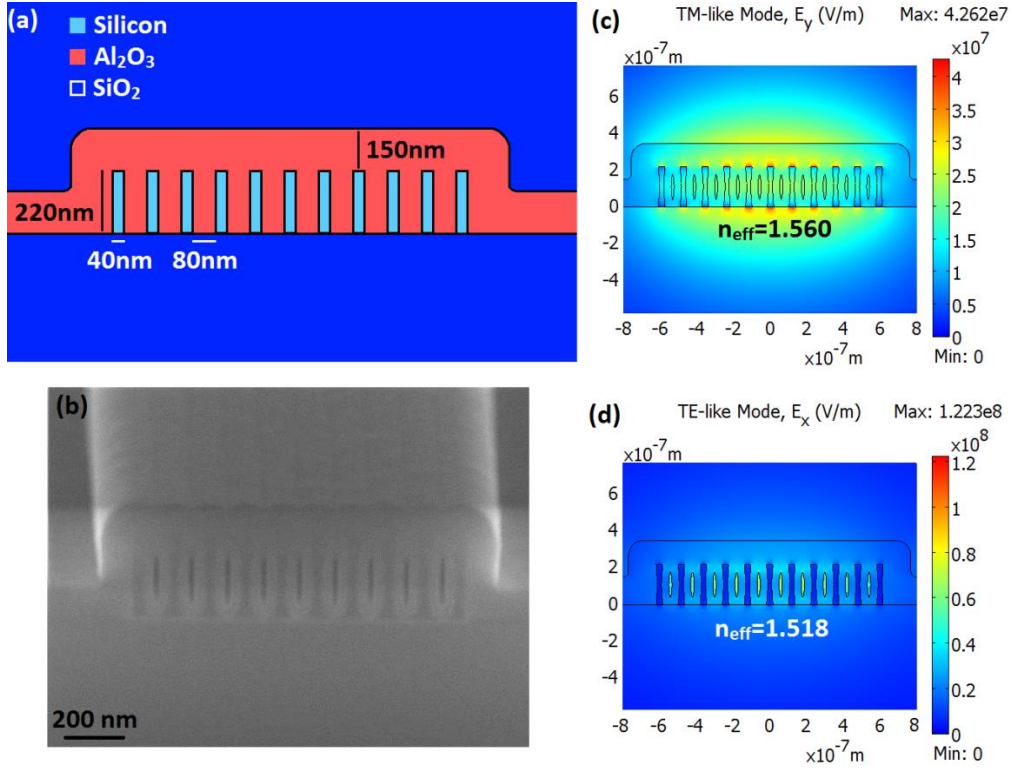

Fig. 4. (a) Target dimensions and (b) SEM micrograph image of the NRA waveguide cross-section. (b,c) FEM-generated plot of the dominant field component for the TM- and TE-like modes, respectively.

## 4. Optical characterization

To measure the linear loss coefficient of the NRA waveguides' TE-like mode, we fabricated an NRA ring resonator coupled to an NRA bus waveguide, as shown in Fig. 5a and Fig. 5b. The separation between the bus and ring waveguides was 100 nm, and the radius of the ring which was experimentally characterized was 40 μm. We coupled light into the bus waveguide using a tunable laser and a lensed tapered fiber, then collected the transmitted power using a metallic output objective and an optical power meter. This measurement setup has been discussed in greater detail in previous work [18]. The normalized transmission spectrum was found to exhibit sharp dips at each of the ring's resonant wavelengths, and the shape of these resonances could be fit numerically to the expression [19]:

$$T = \frac{t^2 - 2t\tau \cos(\theta) + \tau^2}{1 - 2t\tau \cos(\theta) + (t\tau)^2} \quad (1)$$

where θ was the phase accumulated by the mode in one round trip, t was the coupling coefficient from the bus to the ring, τ was the round trip loss within the ring, r was the ring radius, $n_{eff}$ was the effective index of the mode supported by the ring, and λ was the optical wavelength. The loss coefficient could then be calculated as [19]:

$$\alpha\left[\frac{dB}{\mu m}\right] = \frac{-20\log_{10}(\tau)}{L[\mu m]} \quad (2)$$

where L was the length of the ring. The transmission spectrum of the ring resonator, shown in Fig. 5c, displayed resonances with an approximate free spectral range of 5 nm. Our experimental data was used in combination with Eq. 1 and Eq. 2 to calculate both the effective index and loss coefficient of the TE-like mode at a wavelength of 1552 nm, and the two values yielded were, respectively, 1.501±.001 and .0217±.0013 dB/μm. The loss measured here is comparable in magnitude to the values exhibited by subwavelength gratings [15], and it is most likely due both to the NRA waveguide's high surface area to volume ratio and to interface charge-induced free-carrier effects [20]. It is important to note that this value may be significantly reduced in future iterations through the application of a post-etch RCA clean [21].

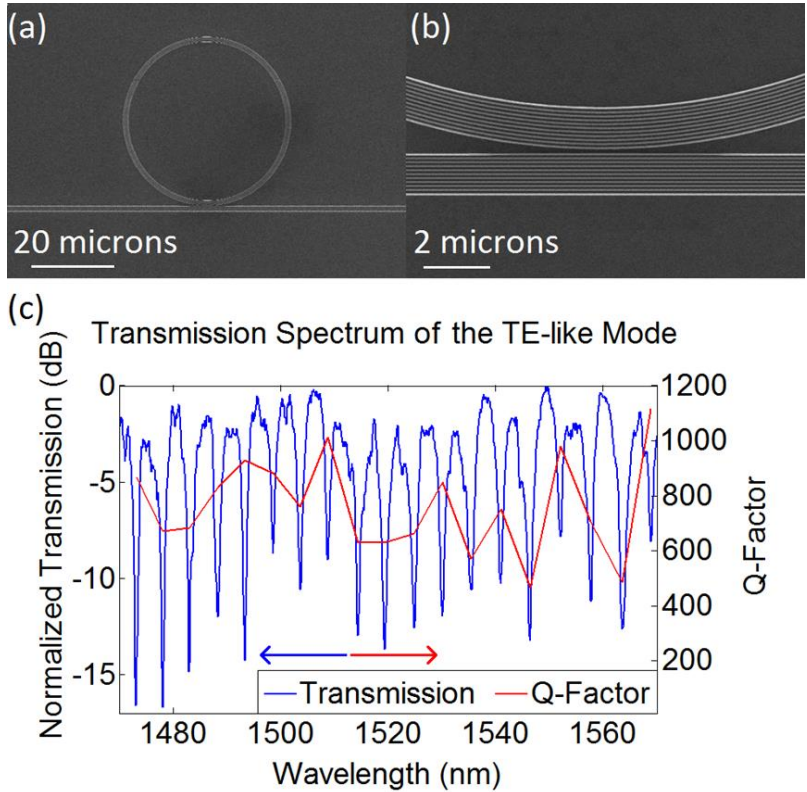

Fig. 5. (a,b) SEM micrographs showing the NRA ring resonator coupled to a bus waveguide, and (c) the transmission spectrum measured at the output of the bus waveguide, additionally showing the Q-factor of each resonance.

## 5. Applications

### 5.1 Nonlinear optics

Materials patterned on a deeply subwavelength scale often exhibit enhancements to the second- and third-order nonlinear susceptibility tensor [22-24], and this effect is believed to be due primarily to (1) the highly localized electric field profiles supported by such structures, and (2) the high spatial density of optical nonlinearities along each material interface. To determine whether such enhancements were present for our fabricated waveguides, we carried out surface third-harmonic generation from large (1 mm by 1 mm) silicon surfaces which were similarly patterned, using a multiphoton microscope as shown in Fig. 6a. The measurements were taken using an unpolarized fs-pulsed laser operating at a wavelength of 1560 nm with a repetition rate of 8 MHz [25]. Fig. 6b shows the multiphoton spectra for the patterned and unpatterned sections of the sample, and this data clearly illustrates that the third-harmonic signal, observed as a sharp peak at 520 nm, increased drastically due to the patterning of the surface. The strongest signal was reached for a ridge width of 70 nm and a ridge periodicity of 70 nm, and the factor of enhancement observed for these dimensions, in comparison to the unpatterned silicon surface, was approximately equal to 500. Because the power contained in the third-harmonic signal was highly dependent on the exact geometry of the patterning, further work may be done to optimize the combined response of the cascaded silicon surfaces and the electric field enhancement. Nonetheless, these results suggest that NRA waveguides may be highly advantageous in processes which rely on third-order parametric wavemixing in integrated devices.

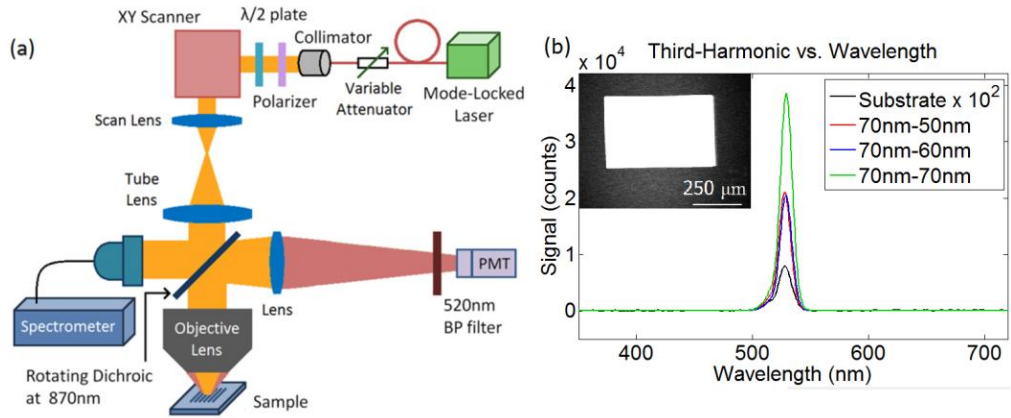

Fig. 6. (a) Optical setup used to measure the third-harmonic signal of the samples. (b) Measured third-harmonic spectra for different silicon surfaces. Inset shows a spatially resolved image of the same signal for one of the patterned regions (inset: spatially resolved image of the THG signal, showing a high degree of enhancement within the patterned region).

*5.2 Mode-split sensing*

As mentioned, NRA waveguides exhibit large spatial discontinuities in their index of refraction. This makes them ideal candidates for sensing applications, particularly if they can be exploited in a resonant configuration. Currently, one of the most common techniques in on-chip nanoparticle detection is to measure the change in the resonant wavelength of a resonator induced by the presence of a nanoparticle. Although effective, this method is limited in terms of the magnitude of the spectral shift it can attain, in addition to being highly susceptible to unintended perturbations to the ring. But perhaps even more importantly, the spectral shifts of such resonators depend strongly on both the nanoparticle's size and location, and decoupling these two variables from one another is not always possible [26,27].

Recently, a self-referencing detection mechanism based instead on the splitting of degenerate modes, achieved using a high-Q silica microtoroid resonator, has been demonstrated to overcome the limitations of conventional chip-scale sensors [28]. This new mechanism relies not on a single mode, but on the degenerate clockwise- and counterclockwise-traveling modes of a perfectly symmetrical ring resonator. Upon the introduction of an impurity, the degeneracy of the two modes is lifted, leading to a split in their resonant frequencies. The strength of this splitting is again strongly dependent on both the size and location of the particle, but in this case both variables may readily be extracted from the resonator's transmission spectrum. Additionally, the two non-degenerate resonances encounter the same noise, and this makes the measurement of their frequency difference more robust than the analogous measurement of traditional sensors.

Here, we evaluate unclad NRA whispering gallery mode (WGM) ring resonators as possible devices for sensing based on self-referential mode-splitting. Because of their topology, trenched structures allow nanoparticles to interact with the core of the optical mode, rather than the evanescent tail. This leads to a stronger perturbation of the structure's resonances, which in turn yields a more appreciable eigenfrequency splitting and a higher overall sensitivity. To demonstrate this effect, two-dimensional models of a 40 μm-radius ring resonator were constructed in COMSOL and used to simulate the eigenmodes supported at 1.55 μm by (1) an unclad NRA waveguide consisting of eleven 50 nm-wide ridges separated by 50 nm, and (2) a traditional ring resonator of comparable size. A dielectric particle with a 30 nm diameter was incorporated into the model to induce the splitting of the resonances' eigenfrequencies, as shown in Fig. 7a-d. To achieve the strongest possible response, the particle was initially assumed to be in the central slot of the NRA waveguide. We observed that the split in resonances, Δf, for the case of the NRA ring was approximately 7 GHz, while the corresponding value of the traditional ring resonator was approximately 8.19 MHz. Fig. 7e illustrates how, even as the particle was moved to slots farther away from the NRA waveguide's center, the splitting of the eigenfrequencies remained orders of magnitude larger than that of the conventional ring.

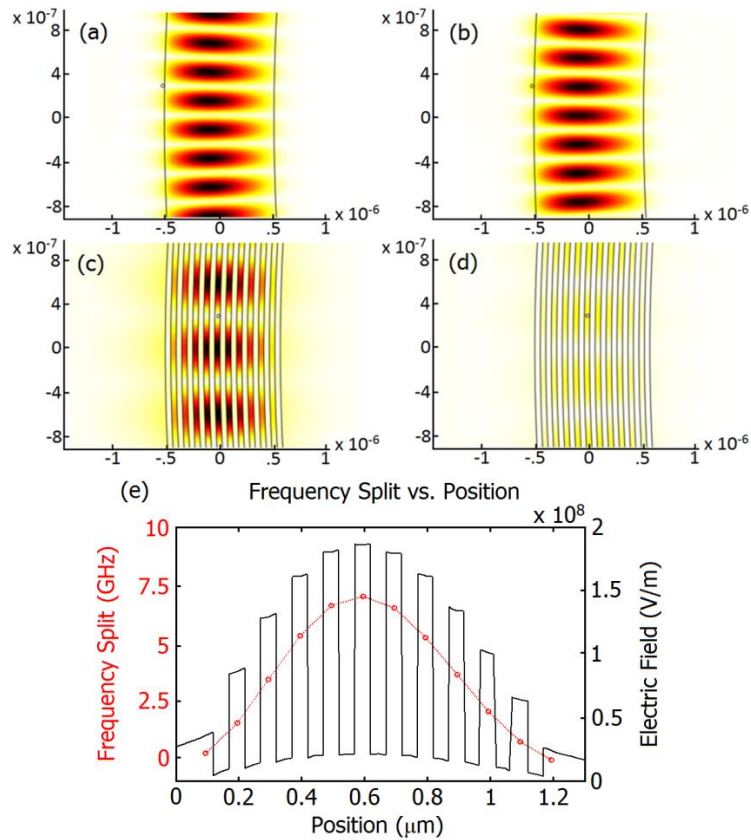

Fig. 7. Non-degenerate modes upon the introduction of a 30 nm-wide dielectric particle in a 40 μm ring resonator for (a,b) a traditional waveguide and (c,d) an unclad NRA waveguide. (e) Frequency splitting of the mode as a function of the particle's position, given in terms of the slot number, in the NRA waveguide.

## 7. Conclusion

To summarize, we have modeled, fabricated, and characterized a new waveguide topology consisting of closely spaced silicon ridges clad with aluminum oxide and silicon dioxide. The waveguide was found to support a TE-like mode which exhibited a competitively low propagation loss coefficient, and was additionally shown to offer unique advantages in terms of both nonlinear optics and sensing applications. As integrated silicon photonics continues to evolve, the NRA waveguide may be leveraged to improve the efficiencies of a wide range of linear and nonlinear optical device components including but not limited to modulators and switches, wave mixers, wavelength- and polarization-sensitive filters, and spectrometers.

## Acknowledgements

This work was supported by the Defense Advanced Research Projects Agency (DARPA), the National Science Foundation (NSF), the NSF ERC CIAN, the Office of Naval Research (ONR), the Multidisciplinary University Research Initiative (MURI), and the Cymer Corporation. Sample fabrication was performed at the UCSD Nano3 cleanroom facility, and sample characterization was performed at the Chip-Scale Photonic Testing Facility, a collaboration between UCSD and Agilent technologies which we gratefully acknowledge.